# Strong field-matching effects in superconducting YBa$_2$Cu$_3$O$_{7-\delta}$ films with vortex energy landscapes engineered via masked ion irradiation

I. Swiecicki[1], C. Ulysse[2], T. Wolf[3], R. Bernard[1], N. Bergeal[3], J. Briatico[1], G. Faini[2], J. Lesueur[3], and Javier E. Villegas[1,*]

[1] *Unité Mixte de Physique CNRS/Thales, 1 avenue A. Fresnel, 91767 Palaiseau, and Université Paris Sud 11, 91405 Orsay, France*

[2] *CNRS, Phynano Team, Laboratoire de Photonique et de Nanostructures, route de Nozay, 91460 Marcoussis, France.*

[3] *LPEM, CNRS-ESPCI, 10 rue Vauquelin 75231 Paris, France*

[*]email: javier.villegas@thalesgroup.com

We have developed a masked ion irradiation technique to engineer the energy landscape for vortices in oxide superconductors. This approach associates the possibility to design the landscape geometry at the nanoscale with the unique capability to adjust depth of the energy wells for vortices. This enabled us to unveil the key role of vortex channeling in modulating the amplitude of the field matching effects with the artificial energy landscape, and to make the latter govern flux dynamics over an usually wide range of temperatures and applied fields.

The energy landscape for vortices in superconductors can be engineered via the introduction of ordered distributions of sub-micrometric structures –e.g. holes or inclusions– which create the energy wells (pinning sites) for vortices. This possibility has opened the door to a wide spectrum of fundamental studies, and to a number of applications (for a review, see Ref. 1). From the fundamental point of view, the interest of flux dynamics on artificial energy landscapes reaches beyond vortex physics, as it has become a model system to study a variety of problems which involve interacting particles (e.g. colloids, atoms, or charge density waves) moving on a pinning substrate[2,3]. Regarding the applications, the ability to manipulate vortices has brought forth the so-called "fluxtronics", in which vortex confinement, guidance[4-7] and rectification[8-11] via energy landscapes with special geometries constitute the basis of superconducting diodes[9-13], signal processing[14] and novel computing applications[15-17].

While much of the progress in this area has been done with low critical-temperature ($T_C$) superconductors[1,2,4,5,8,10,11], extending this research to high-$T_C$ materials[6,7,14,18-22] is particularly interesting. For the latter, the interplay between the artificial ordered pinning, the characteristic anisotropy, and the strong thermal fluctuations predisposes to a richer phenomenology. However, in the presence of strong intrinsic random pinning, the impact of the artificial energy landscape on vortex dynamics –as measured e.g. by the amplitude of the field matching effects observed in the magnetotransport– is generally found to be much weaker in high-$T_C$ (e.g. $YBa_2Cu_3O_{7-\delta}$)[18,21,22] than in low-$T_C$ superconductors (e.g. Nb)[1].

In this paper, we show how to make the artificial energy landscape prevail over the intrinsic random pinning and govern vortex dynamics in $YBa_2Cu_3O_{7-\delta}$ (YBCO) thin films, in an unusually wide range of temperatures, and in relatively high magnetic fields. We use a combination of e-beam lithography and ion irradiation to engineer the vortex energy landscape via the modulation of the local electronic properties produced by ion damage[23]. This form of "electronic patterning" allows shaping the vortex energy landscape at the scale of only a few tenths of nanometers. That is about one order-of-magnitude better than achieved by physically patterning the superconductors via lithography and etching (e.g. introducing arrays of holes)[6,7,9,14,18-20], which dramatically increases –up to the kGauss range– the field range in which the artificial energy landscape is dominant. In addition, and crucially, the method developed here provides with the unique ability to adjust the depth of the landscape energy wells. By systematically varying both parameters (geometry and depth) in a series of experiments, we actually obtained field matching effects which are as strong as for clean low-$T_C$ materials, and we gained understanding on the mechanism determining their amplitude. In particular, our experiments evidence the key role played by vortex channeling effects, which critically depend on the distance between energy wells in the landscape. In addition to its fundamental interest, all of the above makes of our approach a powerful, versatile method for vortex manipulation in oxide superconductors, in applied magnetic fields up to two orders of magnitude higher than with other techniques[6,7,9,14,18-20]. This is specially relevant for fluxtronic devices, because higher applied magnetic fields $B$ imply a higher the vortex densities $n_v \alpha B$, and therefore greater data storage capacity in devices conceived for logic operation[15-17], and larger output signals in those for signal processing[14].

The samples fabrication is done as follows. The YBCO superconducting thin films (50 nm thick; grown by pulsed laser deposition on (100) $SrTiO_3$ substrates) are covered with a thick (~800 nm) resist for e-beam lithography (PMMA). The lithography process allows us to create periodic hole arrays in the PMMA [see Figs. 1 (a) and (b)], with the desired geometry (square, rectangular, etc), distance between holes $d$ (center to center; ranging from 80 to 180 nm), and holes diameter $\varnothing$. The experiments in this paper are for square arrays of holes with fixed $\varnothing=40$ nm and variable $d=120$, 150 and 180 nm. After lithography, the resulting nano-perforated PMMA layer on top of the YBCO layer is used as a mask [Fig. 1 (d)] through which we irradiate with $O^+$ ions (with energy E=110 keV and fluence ranging $f=10^{13}$ cm$^{-2}$ to 5 $10^{13}$ cm$^{-2}$). After irradiation, the resistance *vs.* temperature R(T) of the YBCO films [Fig. 1 (c)] reveal a depression of T$_C$ and an increase of the normal-state resistivity $\rho_N$ as compared to the fresh films. As we show below, these effects depend on the irradiation fluence *f* and on the mask density of holes.

The projected range of penetration of the 110 keV O$^+$ ions into PMMA is ~600 nm (obtained from Monte-Carlo simulations)[24]. Therefore the ions are fully stopped by the mask and reach the YBCO film only through the mask holes. O$^+$ ion bombardment does not change the YBCO surface morphology[21], but creates point defects within the bulk of the material[21]. Since the O$^+$ track length into YBCO (~ 150 nm)[22] is much longer than the film thickness (50 nm), ion-induced damage within the films is expected in depth from the exposed surface, down into the SrTiO$_3$ substrate. The local density of point defects $\sigma$ –defined as the ratio of displaced atoms per existing ones– can be calculated via Monte Carlo simulations[24] which take into account the irradiation energy E, the fluence *f* and the PMMA mask geometry. $\sigma$ (averaged over the film thickness) for different arrays and *f* are shown as contour plots in Figs. 2 (d)-(h). Note that the point defects appear not only underneath the hole areas directly exposed to be ion beam, but also in between these, because the ions spread out as they impinge on the YBCO film. The density of defects in the non-exposed areas of the film strongly depends on the fluence *f* and the distance between mask holes *d*. The presence of point defects implies a local depression of the critical temperature, according to an Abrikosov-Gorkov depairing law[25]. This allows us to calculate the *local* critical temperature t$_C$ expected from $\sigma$. t$_C$(*x*) [with *x* the position along the dashed lines shown in Figs. 2(d)-(f)] is displayed in Figs. (a)-(c) . For the highest *f*=5 $10^{13}$ cm$^{-2}$ [black curves in Figs. 2(a)-(c)], t$_C$ is completely suppressed in the hole areas directly exposed to the ion beam. Lower *f* imply a depressed but finite t$_C$ in the hole areas (see red and green curves in Fig. 2 (a) for 2·$10^{13}$ and $10^{13}$ cm$^{-2}$ respectively). Thus, tuning of *f* and *d* enables us to tailor the spatial modulation of $t_C$. Because it is energetically favourable for flux quanta to locate in regions with depressed $t_C$[26], this allows us to design the vortex energy landscape with nanometric resolution. The potential energy wells for vortices will be formed in areas where $t_C$ is more substantially depressed (i.e. the circular areas exposed to the ion beam).

Fig. 2 shows that the experimental critical temperatures $T_C$ of the films after irradiation (defined by $R(T_C)=0.9R_N$, with $R_N$ the normal-state resistance at the onset of the transition) are in good agreement with the above simulations. The experimental $T_C$ (normalized to the critical temperature prior to irradiation $T_{C\ VIRG}$ ~ 80-85 K for all samples) as a function $d$ and $f$ is shown with solid symbols in Figs. 2 (i) and (j), respectively. Note that in both cases the experimental trends are closely reproduced if one plots as a function of $d$ and $f$ the maximum $t_C$ (which is in the areas non-exposed to the ion beam) obtained from simulations. Finally, we observe that the increase of the normal-state resistivity (at 100 K) with respect to that of the fresh samples ($\rho_0$~500-700 $\mu\Omega$ cm for all samples) scales with $f/d^2$, which means that the films resistivity is proportional to the average irradiation dose.

We show in what follows that the modulation of $t_C$ creates an energy landscape for vortices, and that slight changes of the irradiation fluence and array parameter $d$ produce dramatic effects on flux dynamics. Figs. 3 (a)-(f) depict the resistance *vs.* applied field $B$ for different $d$ and $f$ [each panel correspond to each of the cases for which simulations are shown in Figs. 2 (d)-(h)]. $B$ is applied perpendicular to the film plane. The temperatures for the measurements in Fig. 3(a)-(e) were chosen for all the samples to display similar zero-field resistance (normalized to the normal-state one $R_N$). This criterion enables the direct comparison between the curves, and is more appropriate than the absolute temperature T or the reduced one $T/T_C$, provided the different $T_C$ and transition widths depending on $f$ and $d$. A series of pronounced periodic oscillations are observed in the curves (note that the *y* axis is in logarithmic scale). The amplitude of the magneto-resistance oscillations decreases with increasing spacing $d$ [for fixed $f$, Figs, 3(a)-3c)], and also when the $f$ is decreased [for fixed $d$, Figs. 3(a)-(d)-(e)]. For each curve, the more pronounced minima correspond to the "matching field" $B=\pm B_1$, with $B_1 \equiv \phi_0/d^2$ the field at which the external field induces one flux quantum per unit cell of the square array. This is shown in Fig. 3 (h), in which the experimental $B_1$ are plotted vs. $\phi_0/d^2$. The sample with highest $f=5\ 10^{13}$ cm$^{-2}$ and shortest $d=120$ nm (Fig. 3a and 3f) presents clear minima also at $B=\pm 2B_1$ (two flux quanta per unit cell). However, the second-order minima become less pronounced as $d$ is increased (Figs. 3(a) →(b)→ 3c)). For lower $f$ [Figs. 3(d)-(e)], only first-order matching effects are visible. Note finally that a closer look to the curve in Fig. 3 (f) (see inset) unveils the presence of clear fractional matching at $B=0.5B_1$, for which minima are shallower as expected[27]. In summary, the strongest matching effects (characterized by deeper minima, and the by presence of clearer second order and fractional matching) are observed for the shortest $d$ and highest $f$.

The commensurability effects described above are the well-known fingerprint of periodic flux pinning in superconductors[1,4,5,8,10,14,18-20]. Matching of the flux lattice to the square geometry of the artificial energy landscape produces an enhancement of the vortex-lattice pinning, which slows-down vortex motion and leads to a resistance decrease. In the present experiments, these effects appear in an unusually wide range of temperatures below $T_C$. When considering the magneto-resistance measured at constant injected current, the effects typically smooth out with increasing temperature [see Fig. 3(f)]. In the non-ohmic regimes of the resistance, increasing the injected current also leads to smoothing of the magnetoresistance oscillations. Notably, for the samples irradiated with higher fluences ($f=5\ 10^{13}$ cm$^{-2}$), the matching effects are present both above and below the irreversibility line[24] (which was determined from the curvature of I(V) characteristics). In the latter case, the matching effects are also visible in the critical depinning current $J_C$ vs. field, very far below $T_C$ [see Fig. 3(g)]. This behaviour is unusual in samples with strong random pinning (such as Nb and PLD-grown YBCO thin films), in which that type matching effects are typically found only very close to $T_C$ and gradually disappear for lower temperatures as the disordered pinning prevails over the artificial pinning landscape.[1,18]

The results shown in Fig. 3 illustrate the potential of the irradiation technique to finely tailor the energy landscape for flux quanta. Varying $f$ changes the strength of the commensurability effects [Figs. 3(a)→(d)→(e)], because the amplitude of the $t_C$ modulation diminishes as the irradiation fluence is decreased, which leads to shallower potential energy wells and thus to weaker flux pinning. Interestingly, relatively small variations of the array dimensions also produce dramatic changes on flux dynamics: for fixed $f$, the amplitude of the magneto-resistance oscillations decreases as $d$ is increased [Figs. 3(a)→(c)]. We argue that this behaviour is connected to flux channelling across the energy landscape,[1,5-7] Besides directing vortex motion along preferred directions[5-7] (the energy landscape $x$ and $y$ axes), channelling enhances vortex mobility for fields in which the flux lattice does not match the pinning potential (i.e. for fields different from $B=B_1, 2B_1...$ etc)[1]. Therefore, in the presence of strong vortex channelling, the amplitude of the magneto-resistance oscillations is larger than when channelling is weak, due to a greater difference between the resistances at the matching condition and out-of-matching. Channelling is caused by the overlap of the potential energy wells. The shorter the distance between them, the stronger the channelling effects[1,5-7]. In the present experiments, this effect is exacerbated because the ion damage -and therefore $t_C$- within the areas in between the mask holes strongly depends on the array parameter $d$, as illustrated by Fig. 2. Therefore, when $d$ is the shortest, channelling effects are the strongest, and the magneto-resistance oscillations the largest.

Further evidence for the above scenario is found by comparing the mixed-state magneto-resistance with the magnetic field applied in-plane (i.e. parallel to the *ab* plane; $B_{//}$) and out-of-plane (perpendicular to the *ab* plane, $B_\perp$). This is shown in Fig. 4 for the sample that exhibits the strongest matching effects ($f$=5 10$^{13}$ cm$^{-2}$ and $d$=120 nm). The electrical current is perpendicular to the applied field in both cases.

The inset of Fig. 4 shows the raw R($B$). As expected, the commensurability effects are absent when the applied field is parallel to the film plane: in this case, the Lorentz force is perpendicular to the film plane, and therefore the vortices are forced to move in the direction perpendicular to the periodic energy landscape. The background magneto-resistance is very different depending on the applied field direction. In particular, R($B_{\parallel}$)<<R($B_{\perp}$) as is expected for anisotropic superconductors[26]. We quantitatively analyzed this behaviour via the anisotropic Ginsburg-Landau approximation.[24] In the absence of artificial pinning, this allows scaling the external field via the anisotropy parameter γ and the rule $B=\gamma^{-1}B_{\parallel}=B_{\perp}$, so that R($\gamma^{-1}B_{\parallel}$)=R($B_{\perp}$). For pristine YBCO films γ~5-7. For the irradiated film, however, such scaling is not possible over the entire field range, but only for $B$>~5 kOe. This is shown in the main panel of Fig. 4, which displays the collapse of R($\gamma^{-1}B_{\parallel}$) and R($B_{\perp}$), obtained with γ~4. The comparison of R($\gamma^{-1}B_{\parallel}$) and R($B_{\perp}$) in the field range within which they do not coincide provides with valuable information. Below the first matching field, when the vortex density is low, channelling along the energy landscape principal axes yields a higher vortex mobility for out-of-plane than for in-plane fields; in the latter case there is no channelling since vortices move perpendicularly to the artificial energy landscape. This accounts for R($B_{\perp}$)<R($\gamma^{-1}B_{\parallel}$). For fields closer to the matching condition ($B_{\perp}=B_1$), the pinning enhancement occurring for perpendicular field dramatically reduces the vortex mobility, and therefore R($B_{\perp}$)<<R($\gamma^{-1}B_{\parallel}$). According to previous work[28], R($B_{\perp}$)<<R($\gamma^{-1}B_{\parallel}$) at the matching fields (i.e. for $B=B_1$ and $2B_1$) confirms that the magneto-resistance oscillations are connected to flux pinning phenomena, and rules out that the irradiated film was merely behaving as a superconducting wire network[29] and the oscillatory magneto-resistance originated from Little-Parks[30] or closely related effects[31]. This is further supported by the fact that the periodic field modulation can be seen also in $J_C(B)$ [Fig. 3 (g)], at temperatures far below $T_C$, contrary to what is observed for flux quantization effects in superconducting wire networks[31]. Finally, in the high field regime $B>\sim5$ kOe, even though the commensurability effects are not visible, the presence of the periodic energy landscape produces a somewhat reduced anisotropy ($\gamma\sim4$) as compared to plain YBCO.

In summary, we have a developed a masked irradiation technique to engineer the energy landscape for flux quanta in high-temperature superconductors. This approach offers the unique advantage that it allows i) shaping the vortex energy landscape at a nanometric scale, and ii) independently tuning the depth of the energy wells. Based on this ability, our experiments evidence the role of vortex channelling effects in determining the strength of flied matching effects in artificial energy landscapes. Finally, we demonstrate the possibility to finely control flux dynamics in fields up to two orders of magnitude higher than with conventional lithography/etching techniques. Thus, besides its technological interest, our realization enables access to a interesting a regime in which the flux lattice elastic energy and the artificial pinning energy strongly compete. We stress at this point that, in addition to other oxide superconductors, this masked irradiation technique could be applied to a variety of functional oxides sensitive to local disorder (e.g. ferromagnets or semiconductors)[32] in order to engineer phase segregation at the nanoscale.

Work supported by the French ANR via "SUPERHYBRIDS-II".

**Figure Captions**

**Figure 1 (a)** Scanning electron microscopy of a PMMA mask **(b)** Zoom of the indicated area. **(d)** Resistivity *vs.* temperature of a series of YBCO films, prior (VIRG) and after irradiation through masks with different inter-hole distance *d* (see legend, in nm). Curves normalized to the film resistivity at 100 K prior to irradiation. **(c)** Sketch of the ion irradiation of YBCO through the PMMA mask.

**Figure 2 (a)-(c)** Local critical temperature $t_C$ along the array axes indicated by the dashed lines in (d)-(f). Black curves for $f=5 \; 10^{13}$, green for $f=10^3$ and red for $2 \; 10^{13}$ cm$^{-2}$ **(d)-(h)** Defect density $\sigma$ from Monte-Carlo simulations for the parameters *d* and *f* indicated. The color grade is in logarithmic scale (the legend must be multiplied by $10^{-3}$) and saturates above $5 \; 10^{-3}$ and below $4 \; 10^{-4}$. **(i)-(j)** Superconducting critical temperature after irradiation (normalized to that measured prior to irradiation, $T_{C \; VIRG}$) (i) as a function of the array parameter *d*, (for fixed fluence $f=5 \; 10^{13}$ cm$^{-2}$) and (j) as a function of *f* (for fixed $d=120$ nm). Solid symbols are experimental data and hollow ones correspond to the values expected from simulations. **(k)** Normal-state (at 100 K) resistivity of the irradiated samples, normalized to that measured prior to irradiation ($\rho_0$) as a function $f/d^2$. The lines are a guide to the eye.

**Figure 3 (a)-(e)** Mixed-state resistance (normalized to the normal-state one $R_N$) as a function of the applied field *B*, different *d* and *f* (see legends). The injected current was $J=0.5$ kA cm$^{-2}$. The vertical dashed lines mark the first and second order matching fields. **(f)** Same as in (a) at two different temperatures and $J=2.5$ kA cm$^{-2}$. $B_1=1.45$ kGauss is the matching field. Inset: zoom of the curve. **(g)** Critical current as a function of the applied field for the same sample as in (a) at T=12K=0.19T$_C$. **(h)** Experimental matching fields $B_1$ as a function of $\phi_0/d^2$.

**Figure 4:** Magneto-resistance with the field applied perpendicular ($B_\perp$) and parallel ($B_\parallel$) to the *ab* plane, for the sample with $f=5 \; 10^{13}$ cm$^{-2}$ and d=120 nm, at T=27 K and with J=1.25 kA

cm$^{-2}$. The inset shows the raw data and the main panel the curves collapse using the anisotropy γ~4.

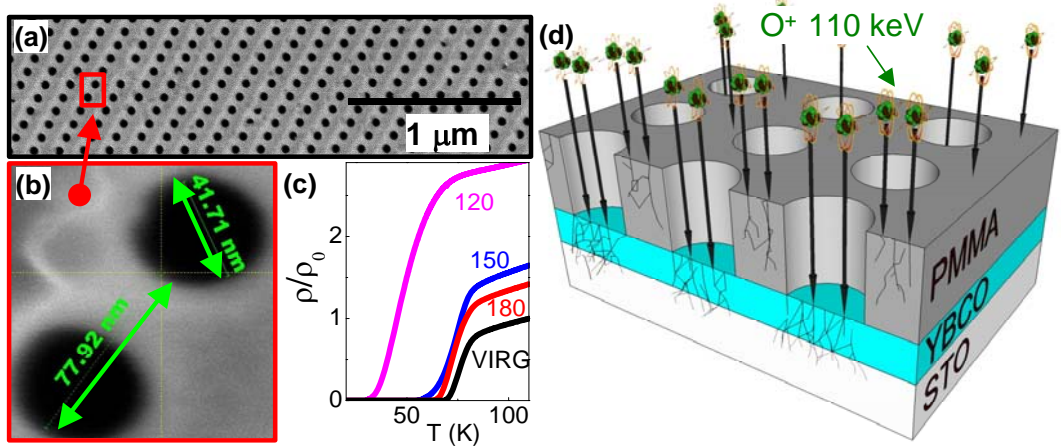

Figure 1

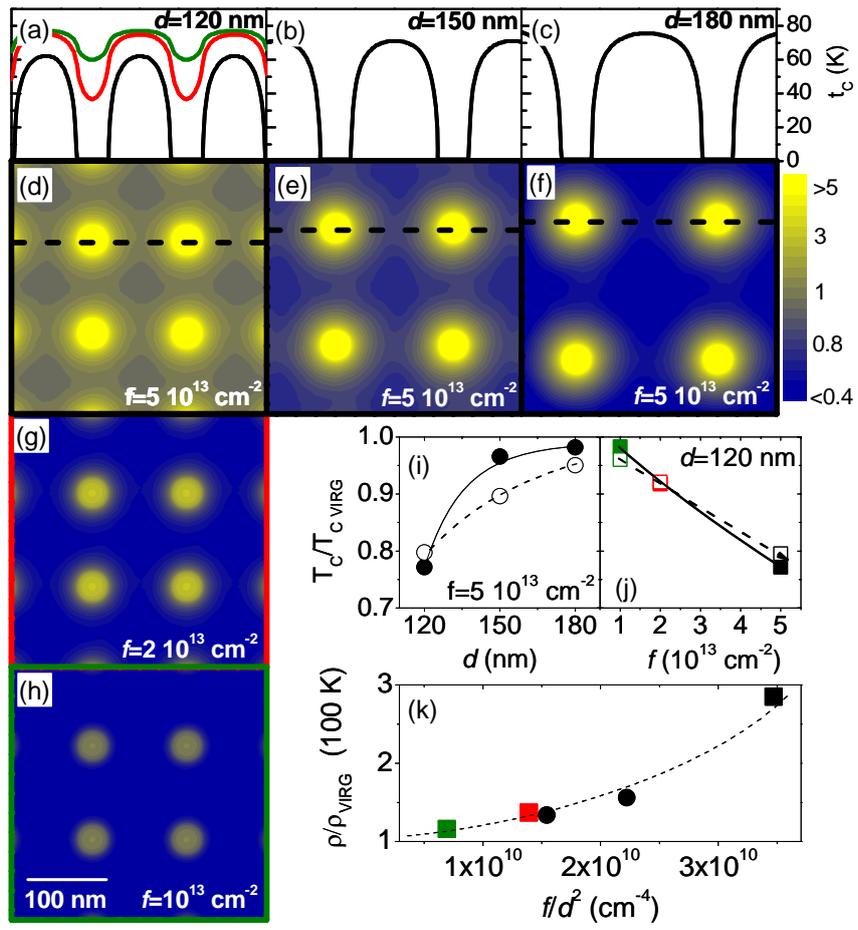

Figure 2

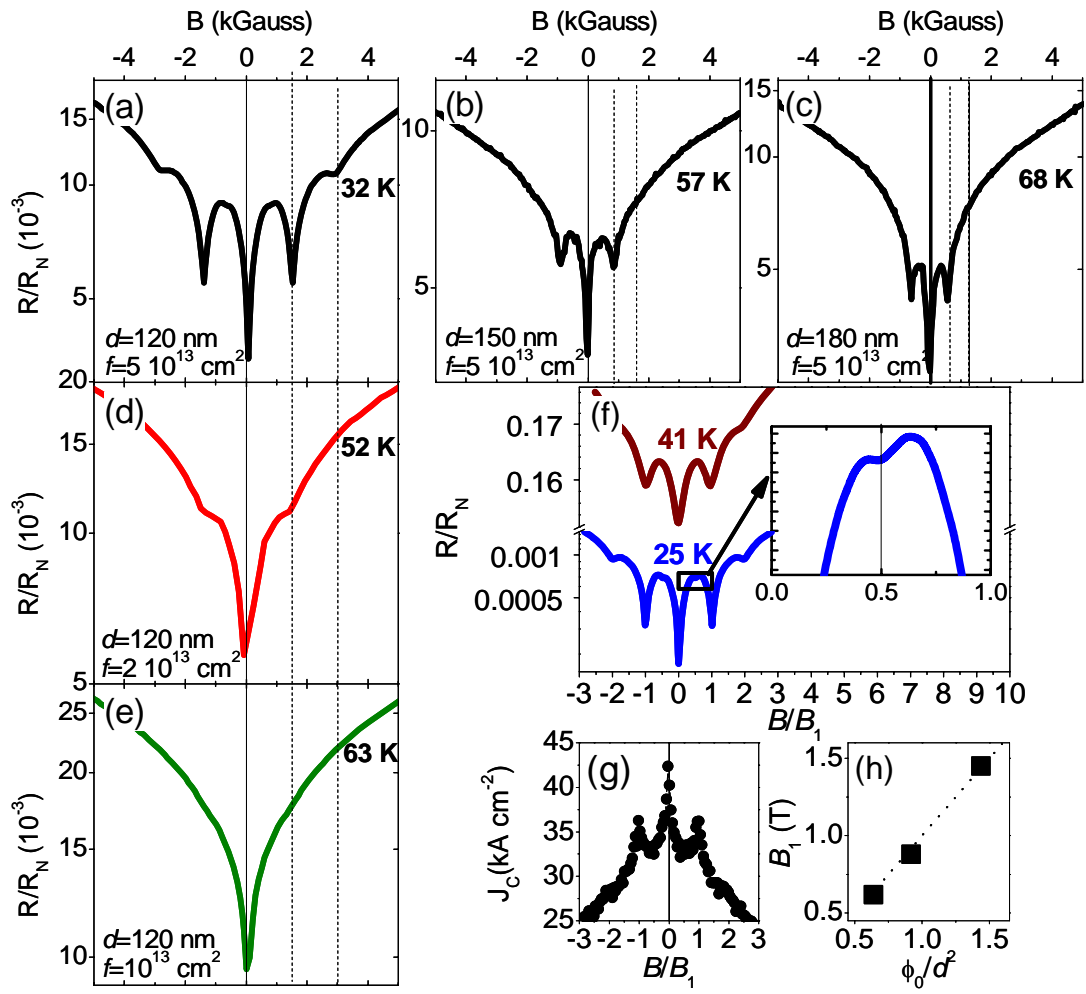

Figure 3

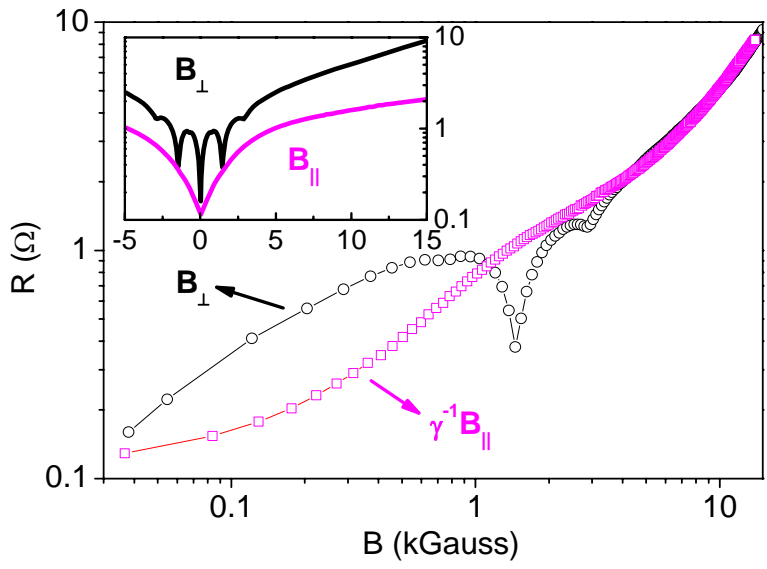

Figure 4